# Application of WLS strips for position determination in Strip PET tomograph based on plastic scintillators

J. Smyrski[1], P. Moskal[1], T. Bednarski[1], P. Białas[1], E. Czerwiński[1], Ł. Kapłon[1,2], A. Kochanowski[2], G. Korcyl[1], J. Kowal[1], P. Kowalski[3], T. Kozik[1], W. Krzemień[1], M. Molenda[2], Sz. Niedźwiecki[1], M. Pałka[1], M. Pawlik[1], L. Raczyński[3], Z. Rudy[1], P. Salabura[1], N.G. Sharma[1], M. Silarski[1], A. Słomski[1], A. Strzelecki[1], W.Wiślicki[3], M. Zieliński[1], N. Zoń[1]

[1]Institute of Physics, Jagiellonian University, Cracow, Poland
[2]Faculty of Chemistry, Jagiellonian University, Cracow, Poland
[3]Świerk Computing Centre, National Centre for Nuclear Research, Otwock-Świerk, Poland

**Abstract:** A method of determination of a gamma quantum absorption point in a plastic scintillator block using a matrix of wavelength-shifting (WLS) strips is proposed. Application of this method for improvement of position resolution in newly proposed PET detectors based on plastic scintillators is presented. The method enables to reduce parallax errors in reconstruction of images which occurs in the presently used Positron Emission Tomography scanners.

## Section 1: Introduction

Plastic scintillators are characterized by relatively short light pulses with decay time on the order of a nanosecond and, therefore, they are widely used in nuclear and particle physics experiments for fast timing measurements. Typically, they have a form of a strip with rectangular cross-section and are read out at both ends by photomultipliers (see e.g. Ref.[1]). Also other solutions are used such as for example scintillator plates read out by arrays of photomultipliers [2,3].

The high timing resolution offered by the plastic scintillators is exploited in a newly invented type of positron emission tomograph (PET) using such scintillators for detection of the 511 keV gamma quanta originating from positron annihilation. Two alternative solutions of the tomograph were proposed [4]. One solution, referred to as the strip PET [5], contains scintillator strips read out by pairs of photomultipliers and arranged around a cylindrical surface forming a tomograph tunnel. Position of the gamma quantum interaction point in the strip - further on we call it shortly "the interaction point" - is determined on the basis of a time difference in propagation of light pulses registered by the pair of photomultipliers.

The second solution, referred to as the matrix PET [6], uses plastic scintillator plates read out by arrays of photomultipliers. Registered amplitude and time of propagation of light pulses allow for localization of the interaction point in the plate. A key feature of both solutions is a high precision in measurement of difference in a time-of-flight (TOF) of the annihilation quanta allowing for determination of position of the positron annihilation along a line of response (LOR). This feature allows for substantial suppression of background in the reconstructed PET images and is one of the main advantages of the plastic scintillators compared to essentially slower inorganic crystals which are used in the contemporary commercial PET scanners.

A disadvantage of the plastic scintillators compared with the inorganic crystals is substantially lower detection efficiency for the gamma quanta. It can be compensated by increasing a length and a thickness of the scintillator segments. However, it leads to worsening of the timing resolution and thus also of the position resolution. Even if a very high precision for measurement of the time difference of 100 ps (FWHM) is assumed, the resulting position resolution is only moderate and equals 7.5 mm (FWHM). This estimation was done taking into account that a speed of propagation of light in a scintillator strip is roughly two times smaller than in vacuum. We propose to improve the position resolution by additional detection of scintillation light escaping the scintillator segments with matrixes of WLS strips.

Read out of plastic scintillators by means of WLS elements is a well established techniques which is applied in many particle physics detectors [7]. Usage of the WLS strips was also proposed for read out of arrays of inorganic crystals in PET detectors [8].

## Section 2: Position determination by means of WLS strips

For determination of a position of the interaction point in a plastic scintillator we propose to use a set of parallel WLS strips which register scintillation photons escaping the scintillator. This concept is illustrated in Fig. 1 showing a side view of a scintillator strip and a set of parallel WLS strips placed above the scintillator. In the figure, we introduced y-z coordinate system with the origin (y=z=0) located in a geometrical center of the scintillator strip and the z-axis oriented along the strip.

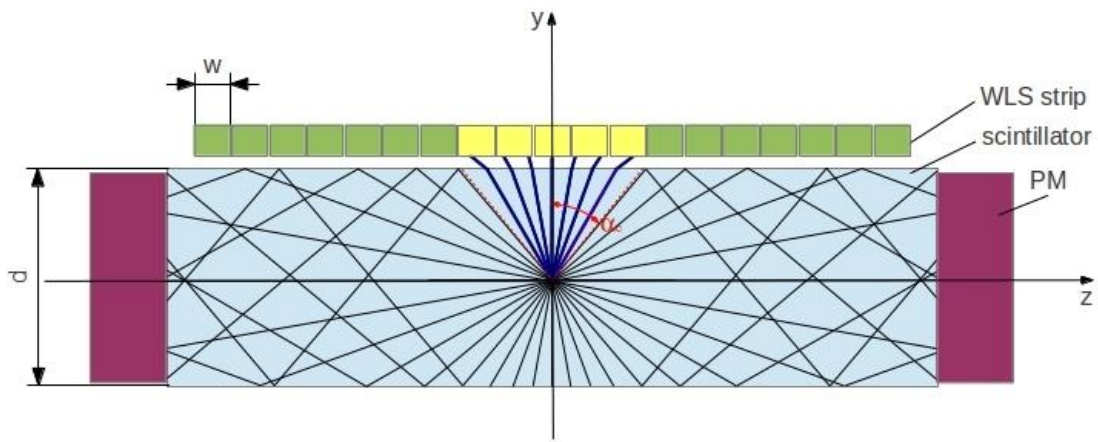

**Fig. 1.** Schematic drawing illustrating application of WLS strips for determination of coordinates of the interaction point in a scintillator strip.

For illustration of propagation of the scintillation photons in the strip, indicated are trajectories of photons emitted from the geometrical center at every 10º with respect to the y-axis. For emission angles larger than the critical angle, which for plastic scintillator material with refractive index n=1.58 equals to 39.2º, the emitted photons undergo total internal reflections from the walls of the scintillator strip and propagate towards the photomultipliers (PMs). Their trajectories are indicated with thin, black lines. For emission angles smaller than the critical angle, scintillation photons can escape the scintillator strip through a side wall and are absorbed in the WLS strips. Trajectories of such photons are indicated as thick, blue lines and the WLS strips which absorb these photons are marked with yellow color. Applied WLS material is selected in such a way, that it absorbs photons with wavelength range characteristic for emission spectrum of the scintillator (see Fig.2). Secondary photons emitted isotropically by the

WLS strips propagate towards photomultipliers attached at their both ends. The isotropic emission of photons with the wavelength corresponding to small absorption in the WLS constitutes the crucial feature of the presented solution, and it causes that some of the secondary photons are trapped in the WLS fiber and propagates via internal reflections towards its edges.

Coordinate of the interaction point along the scintillator strip (z-axis) is determined as a weighted average of z-coordinates of WLS strips with weights equal to amplitudes of signals registered in the WLS strips being proportional to the numbers of absorbed scintillation photons. We expect, that with a width of the WLS strip of w = 5 mm the resolution of the z coordinate should be of about 5 mm (FWHM) and thus it will be better than the one derived from the time difference measured at the strips ends. However, the time information from the photomultipliers reading out the scintillator strips remains important for the TOF measurement of the annihilation gamma quanta.

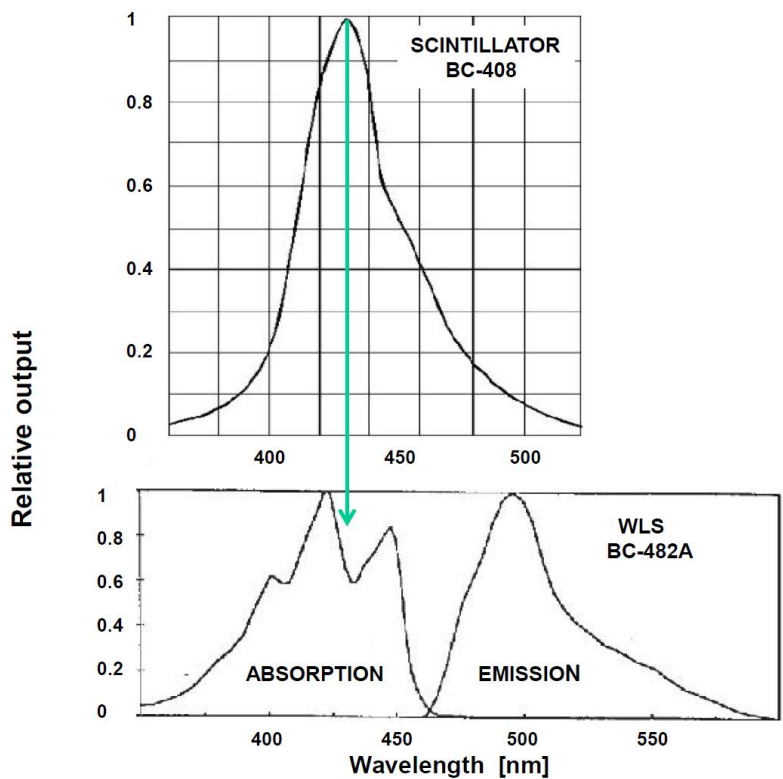

**Fig. 2.** Emission and absorption spectra for the exemplary plastic scintillator and wavelength shifter. The figures have been adapted from Ref. [9].

Position of the interaction point along the y-axis is determined from the number of WLS strips which registered the photons. For emission from the center of the scintillator strip (y = 0) with a scintillator thickness of d=30 mm, and width of WLS strips of w = 5 mm, scintillation photons reach on average 5 WLS strips as shown in Fig. 1. For scintillations occurring close to the scintillator surface adjacent to the WLS strips (y = +d/2), signal is measured only in one or two WLS strips. For emission occurring near the opposite scintillator surface (y = −d/2) the photons reach around 10 WLS strips. The above examples indicate that in principle with WLS strips of 5 mm width a resolution of about 5 mm may also be achieved for the determination of y coordinate of the interaction point. For example one may use a following formula: $DOI \equiv y = d * (n_{max} - 2n) / 2n_{max}$, where n denotes the number of WLS strips which gave a signal, and $n_{max}$ stands for a maximum value of strips which can see photons produced in a single interaction. In the example shown in Fig 1. $n_{max}$=10. Measurement of the y-coordinate is of high importance in PET scanners since it allows to correct the LOR for a depth of interaction (DOI) and thus to avoid parallax errors.

## Section 3: Application of WLS strips in PET detectors

For application of the WLS read out in the strip PET, the scintillator strips have to be grouped in planar segments. Each segment is read out by a set of WLS strips which are oriented at a right angle with respect to the scintillator strips. A schematic drawing of a segment containing 26 scintillator strips and 35 WLS strips is shown in Fig. 3.a. With a chosen width of the scintillator and WLS strips of 5 mm, the segment covers an active area of 130 mm x 175 mm. Coordinate of the interaction point along the scintillator strip (z-axis in Fig. 3.a) and in direction perpendicular to the detection segment (y-axis) is derived from signals of the WLS strips as described in the previous section. The x-coordinate is determined by identification of the scintillator strip that registered the gamma quantum. For detection of light emitted by the scintillator and WLS strips, instead of traditional vacuum photomultipliers, the silicon photomultipliers (SiPMs) can be applied. Due to very small sizes, the SiPMs allow to construct compact detection segments and thanks to a capability to work in high magnetic fields, they can be used in a combined PET/NMR scanner.

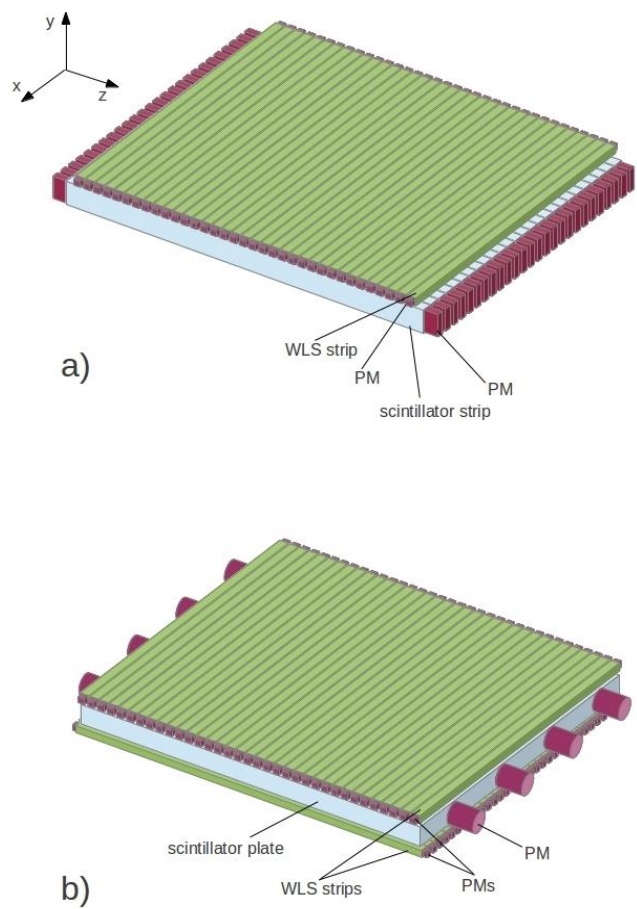

**Fig. 3.** Arrangement of WLS strips for position reconstruction in a set of parallel scintillator strips (a) and in a scintillator plate (b).

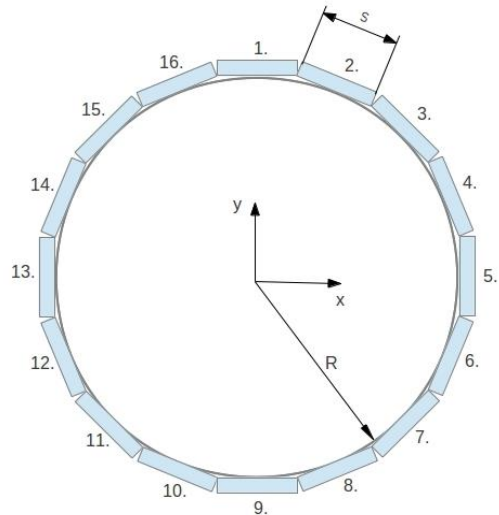

**Fig. 4.** Arrangement of 16 detection segments around a tunnel of PET tomograph.

An example of an arrangement of 16 detection segments in a PET scanner is shown schematically in Fig. 4. With a width of the segments s=130 mm, a cylindrical surface representing a scanner tunnel has a radius of R=327 mm.

In the matrix PET [6], arrays of photomultipliers are replaced by two planes of WLS strips. The planes are parallel to each other, while WLS strips from one plane are perpendicular to WLS strips of the second plane (see Fig. 3.b). This arrangement allows for a three dimensional reconstruction of the interaction point. Additionally, for the TOF measurements, a set of photomultipliers is attached to side walls of the scintillator plate.

## Section 4. Conclusions

Application of the WLS strips in the strip PET and in the matrix PET has been proposed for a three dimensional reconstruction of the gamma quantum interaction point. In particular, it allows to determine a depth of interaction which is important for elimination of parallax errors in reconstruction of images which occurs in the present PET scanners. With a choice of a width of the scintillator strips and of the WLS strips of around 5 mm, a position resolution on this level is feasible and thus is comparable with precision of PET scanners based on blocks of inorganic crystals.

A prototype PET scanner based on plastic scintillators is being constructed in the Institute of Physics of the Jagiellonian University in Krakow. Studies of application of WLS strips for improvement of the position resolution in the scanner are also going on.

## Acknowledgements

We acknowledge technical and administrative support by M. Adamczyk, T. Gucwa-Rys, A. Heczko, M. Kajetanowicz, G. Konopka-Cupiał, J. Majewski, W. Migdał, A. Misiak, and the financial support by the Polish National Center for Development and Research through grant INNOTECH-K1/IN1/64/159174/NCBR/12, the Foundation for Polish Science through MPD programme and the EU and MSHE Grant No. POIG.02.03.00-161 00-013/09.